\documentclass[aps,prl,twocolumn,showpacs,floatfix]{revtex4}

\usepackage{amsmath}
\usepackage{graphicx}
\usepackage{latexsym}
\usepackage{color}

\newcommand{\be}{\begin{equation}}
\newcommand{\ee}{\end{equation}}
\newcommand{\bea}{\begin{eqnarray}}
\newcommand{\eea}{\end{eqnarray}}

\begin{document}

\title{Comment on ``Growth Inside a Corner: The Limiting Interface Shape"}
\author{Rajeev Singh}
\email{rajeev@imsc.res.in}
\affiliation{The Institute of Mathematical Sciences, CIT Campus, Taramani, 
Chennai-600113, India}
\author{R. Rajesh}
\affiliation{The Institute of Mathematical Sciences, CIT Campus, Taramani, 
Chennai-600113, India}
\date{\today}

\maketitle

In a recent letter, Olejarz et. al.\cite{olejarz2012} conjecture the 
asymptotic shape of a crystal, grown by depositing cubes inside a three 
dimensional corner, by generalizing the known two dimensional results 
consistent with the symmetries of the problem. The conjecture differs 
from numerical simulations by $0.9\%$, but the discrepancy is ascribed to 
a slow approach to the asymptotic answer. We do Monte Carlo simulations 
which avoid these transients  and conclude that the conjecture is
inconsistent with our numerical results.

The growth model in Ref.~\cite{olejarz2012} is a solid on solid (SOS) 
model where the integer height $z(x,y)$ at a lattice point $(x,y)$ is 
constrained by $z(x,y) \leq \min[z(x-1,y), z(x,y-1)]$, with the boundary 
conditions $z(x,y) =\infty$ if $x<0$ or $y<0$. At time $t=0$, $z(x,y)=0$ 
for $x,y \geq 0$. With rate $1$, $z(x,y)$ increases by $1$ provided the 
new configuration is a valid one. The conjecture for the asymptotic 
shape, when projected along the $(1/\sqrt{3}, 1/\sqrt{3},1/\sqrt{3})$
direction, reduces to $x=y=z=w t$, where $w=1/8$.
\begin{figure}[h!]
\includegraphics[width=5.1cm]{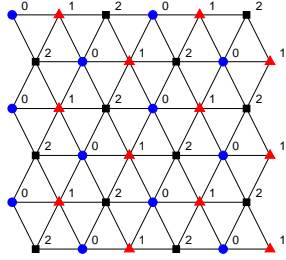}
\caption{On projecting the surface of the crystal onto the $(1,1,1)$
plane, a triangular lattice is obtained. 
If $h_i$ is the height at a site on sublattice $i$, 
then $h_i~ \rm{mod}~ 3 =i$. In addition, $|h_i - h_j|<3$ for all nearest 
neighbor pairs $\langle ij \rangle$.
\label{fig:fig1}
}
\end{figure}

The growth velocity in any direction is a function of the local slope 
$\partial z /\partial x$ and $\partial z/ \partial y$, and is best 
determined using a different set of boundary conditions when these 
slopes are uniform everywhere \cite{rajesh1998}. Thus, for measuring 
$w$, we project the growing crystal onto the $(1,1,1)$ plane to obtain a 
restricted solid on solid (RSOS) model on a triangular lattice, to which 
we apply periodic boundary conditions. The heights on sublattice $i$ of 
the triangular lattice (see Fig.~\ref{fig:fig1}) is $i$ modulo 3. In 
addition, there is a constraint $|h_m-h_n |<3$ for nearest neighbor 
sites $m$, $n$. With rate $1$, the height at a site increases by $3$ 
provided the new configuration satisfies the RSOS condition. The 
periodic boundary conditions restores translational invariance, and we 
measure in the steady state the fraction of sites that can increase in 
height. It is straightforward to see that this fraction is identical to $w$.

\begin{figure}[h!] 
\includegraphics[width=\columnwidth]{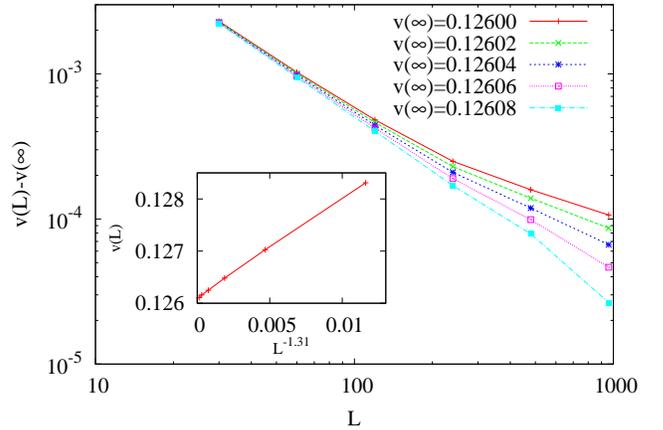} 
\caption{The difference 
between the velocity $w(L)$ and the asymptotic result $w(\infty)$ as a 
function of system size $L$ for different $w(\infty)$. The curve is
roughly a straight line for $w(\infty) = 0.126065$. For smaller (larger) 
$w(\infty)$, the data curve upwards (downwards).  Inset: $w(L)$ against 
$L^{-1.25}$ is a straight line with intercept larger than $0.126$. 
\label{fig:fig2}
}
\end{figure}

We consider systems of size $L\times L$ where $L=30 \times 2^n$, with 
$n=0,\ldots,5$. The steady state averaged data for $w$ is shown in 
Fig.~\ref{fig:fig2}, where the error in each data point is  estimated
by doing many independent runs. We estimate $w=0.12606 (2)$, different from the 
conjectured value $w=1/8=0.125$. While our
estimate for $w$ is consistent with the simulation results in 
Ref.~\cite{olejarz2012}, the discrepancy from $1/8$ can no longer be 
ascribed to transients.
We also observe that $w$ reaches its 
asymptotic value as $L^{-\theta}$ with $\theta \approx 1.25$ (see inset
of Fig.~\ref{fig:fig2}). 

It may be that the asymptotic shape is a linear combination of the two
solutions outlined in Ref.~\cite{olejarz2012}. If that is the case, the relative
weights have to be shown to be independent of $\partial z
/\partial x$ and $\partial z/ \partial y$. This is again best
demonstrated
on a triangular lattice with appropriate boundary conditions.

\end{document}